\documentclass[aps,showpacs,preprintnumbers,amsmath, amssymb]{revtex4}

\usepackage{float}
\usepackage{graphics,epsfig}
\usepackage{graphicx}
\usepackage{dcolumn}
\usepackage{bm}

\begin{document}
\baselineskip=0.8 cm
\title{{\bf Dynamical perturbations and critical phenomena in \\
 Gauss-Bonnet-AdS black holes}}

\author{Yunqi Liu$^{1}$, Qiyuan Pan$^{1}$, Bin Wang$^{2}$, Rong-Gen Cai$^{3}$}
\affiliation{$^{1}$ Department of Physics, Fudan University, Shanghai 200433, China} \affiliation{$^{2}$ Department of Physics, Shanghai Jiao
Tong University, Shanghai 200240, China} \affiliation{$^{3}$ Institute of Theoretical Physics, Chinese Academy of Sciences, P.O. Box 2735,
Beijing 100190, China}

\vspace*{0.2cm}
\begin{abstract}
\baselineskip=0.6 cm
\begin{center}
{\bf Abstract}
\end{center}

We investigate the perturbations of charged scalar field in
$5$-dimensional Gauss-Bonnet AdS black hole backgrounds. From the
perturbation behaviors we obtain the objective picture on how the
high curvature influence the spacetime perturbation and the
condensation of the scalar hair. The high curvature effects can also
be read from the linear response function such as the susceptibility
and the correlation length, when the system approaches the critical
point. We find that the Gauss-Bonnet term does not affect the
critical exponents of the system and they still take the mean-field
values.
\end{abstract}

\maketitle
\newpage
\vspace*{0.2cm}

\section{Introduction}

The holographic model of superconductors, which is constructed by a
gravitational theory of a Maxwell field coupled to a charged complex
scalar field via anti-de Sitter/conformal field theory (AdS/CFT)
correspondence \cite{Witten,Maldacena-1,Maldacena-2}, has been
investigated extensively in recent years (for reviews, see Refs.
\cite{HartnollRev,HerzogRev,HorowitzRev}). According to the AdS/CFT
dictionary, the emergence of the scalar hair in the bulk AdS black
hole corresponds to the formation of a charged condensation in the
boundary dual CFTs. This brings a remarkable connection between the
condensed matter and the gravitational physics which attracts
considerable interest for its potential applications to the
condensed matter physics \cite{HorowitzPRD78}-\cite{Brihaye}. At the
moment when the condensation occurs in the boundary CFT and in the
gravitational counterpart a non-trivial hair for the black hole is
triggered, there appears a phase transition \cite{gubser,Hart}. The
phenomenological signature of this phase transition was recently
disclosed in the perturbation around such AdS black holes
\cite{HeXi,ZhangCai}.

Motivated by the application of the Mermin-Wagner theorem to the
holographic superconductors there were studies of the effects of the
curvature corrections on the (3 + 1)-dimensional superconductor
\cite{Gregory,Pan-Wang,Ge-Wang,Pan2,Brihaye}. It was found that
higher curvature corrections make condensation harder. In addition,
the large Gauss-Bonnet factor gives the correction to the disclosed
universal value for the conductivity $\omega_g/T_c\approx 8$
\cite{HorowitzRev} in the probe limit
\cite{Gregory,Pan-Wang,Ge-Wang}. Furthermore Brihaye \emph{et al.}
observed that the decrease of the critical temperature at which
condensation sets in is stronger as the Gauss-Bonnet coupling
increases, which happens  even beyond the probe approximation
\cite{Brihaye}. In order to get more objective picture on the
influence given by the high curvature on the condensation, in this
work we are going to study the perturbation in the $5$-dimensional
Gauss-Bonnet-AdS black hole backgrounds. We will concentrate on the
bulk high temperature AdS black holes and pay more attention on how
the Gauss-Bonnet term  influences the perturbation in the bulk
background spacetime when the temperature of the black hole drops.
Further we are going to study the critical phenomenon once the AdS
black hole approaches marginally stable mode and the charged scalar
field starts to condensate. We will examine how the Gauss-Bonnet
term affects the critical behavior. Recently, Maeda \emph{et al.}
\cite{maeda} studied most of the static critical exponents of
holographic superconductors for a Reissner-Nordstr$\ddot{o}$m (RN)
AdS black hole with planar horizon and found that they take the
standard mean-field values. We will generalize their work to the
$5$-dimensional Gauss-Bonnet-AdS black hole configurations and
examine the effect of Gauss-Bonnet term on the critical behavior. We
will focus our attention on the high-temperature phase for
simplicity and study the linear perturbations of the bulk equations
of motion in the probe approximation.

The outline of this work is as follows. In section II, we deal with
the perturbation equations of the charged scalar field in the
$5$-dimensional Gauss-Bonnet-AdS black hole spacetime. In section
III, we investigate the perturbation in the bulk and examine the
critical phenomenon of the holographic superconductors. We will
conclude in the last section with our main results.

\section{Perturbation Equations in the Gauss-Bonnet AdS Black Hole}

Let us begin with the $D=p+2$ dimensional charged Gauss-Bonnet black
hole described by the metric~\cite{BD,Cai,Cai2,Cvet,ge}
\begin{eqnarray}\label{metric}
ds^{2}&=&-H(r)dt^{2}+H^{-1}(r)dr^{2}+\frac{r^{2}}{l^{2}}dx_{p}^{2}~,
\end{eqnarray}
where the $U(1)$ gauge field reads
\begin{eqnarray}\label{potential}
A_{t}&=&\frac{Q}{4\pi(D-3)}~(r_{H}^{D-3}-r^{D-3}).
\end{eqnarray}
Here
\begin{eqnarray}
H(r)&=&\frac{r^{2}}{2\alpha}\left[1-\sqrt{1-\frac{4\alpha}{l^{2}}\left(1-\frac{m
l^{2}}{r^{D-1}}+\frac{Q_{0}^{2}l^{2}}{r^{2D-4}}\right)}\right]~,
\end{eqnarray}
where $\alpha$ is the Gauss-Bonnet coefficient, $r_{H}$ is the
horizon radius and $l$ corresponds to the AdS radius. The
gravitational mass $M$ and the charge $Q$ are expressed as
\begin{eqnarray}
M&=&\frac{(D-2)mV_p}{16\pi G_{D}}~,\nonumber\\
Q^{2}&=&\frac{2\pi (D-2)(D-3)Q_{0}^{2}}{G_{D}}~,\nonumber
\end{eqnarray}
where $V_p$ is the volume of the $p$-dimensional Euclidean space
$dx_p^2$ and $G_D$ is the $D$-dimensional Newton constant. Note that
in the asymptotic region (r$\rightarrow\infty$), we have
\begin{equation}
H(r)=\frac{r^{2}}{2\alpha}\left(1-\sqrt{1-\frac{4\alpha}{l^{2}}}\right)~.\nonumber
\end{equation}
We can define the effective  AdS radius
\begin{eqnarray}
\quad\quad\quad\quad\quad
l_{eff}^{2}=\frac{2\alpha}{1-\sqrt{1-\frac{4\alpha}{l^2}}}\sim\left\{
\begin{array}{ll} l^2~, &
~~~~\alpha\rightarrow 0~, \\
\frac{l^2}{2}~, & ~~~~ \alpha\rightarrow \frac{l^2}{4}~.
\end{array} \right.
\end{eqnarray}
The upper bound of the Gauss-Bonnet coefficient $\alpha\leq l^2/4$
is known as the Chern-Simons limit. Besides there also exists a
lower bound $\alpha\geq-\frac{(3D-1)(D-3)}{4(D+1)^{2}}$ by
considering the causality of dual field theory on the boundary
\cite{ge,Buchel,Boer}. Using a coordinate transformation, the metric
(\ref{metric}) and the potential (\ref{potential}) can be rewritten
as
\begin{equation}\label{new metric}
ds^{2}=\frac{l^{2}}{u^{2}}\left[-\frac{r_{H}^{2}(1+c)^{\frac{2}{1-p}}f(u)}{J^{2}(u)}dt^{2}+\frac{J^{\frac{2}{p-1}}(u)}{f(u)}du^{2}
+r_{H}^{2}(1+c)^{\frac{2}{1-p}}J^{\frac{2}{p-1}}(u)dx_{p}^{2}\right]~,
\end{equation}
\begin{eqnarray}
A_{t}&=&\mu \left[1-\frac{1+c}{J(u)}u^{p-1}\right]~,
\end{eqnarray}
with
\begin{eqnarray}
Q_{0}&=&c^{\frac{1}{2}}r_{H}^{p-2}~,\nonumber\\
u&=&\frac{r_{H}}{r(1+c-c~r_{H}/r)^{p-1}}~,\nonumber\\
J(u)&=&1+cu^{p-1}~,\nonumber\\
g(u)&=&J^{\frac{2p}{p-1}}(u)-(1+c)^{\frac{2p}{p-1}}u^{p+1}~,\nonumber\\
f(u)&=&\frac{J(u)}{2\alpha}\left[J(u)-\sqrt{J^{2}(u)-4\alpha
g(u)}\right]~,
\end{eqnarray}
where $\mu$ is the chemical potential and $c$ is related to the
parameter $Q_{0}$. Obviously $u=0$ is the AdS boundary and $u=1$ is
the location of the horizon. The metric (\ref{new metric}) will be
reduced to the $D$-dimensional RN-AdS black hole if we take the
limit $\alpha\rightarrow0$. On the other hand, it becomes the
neutral Gauss-Bonnet-AdS black hole~\cite{Cai} if $Q_0 = 0$. There
are four parameters $\alpha,~c,~r_{H}~$and $\mu$ which parameterize
the background (\ref{new metric}). In fact, not all of them are
independent, they are related by
\begin{equation}
\rho=l^{p-2}r_{H}^{p-1}~\mu~,
\end{equation}
where $\rho$ is the charge density. The temperature $T$ and the
chemical potential $\mu$ of the  black hole are given by
\begin{eqnarray}
T=\frac{p+1-(p-1)c~r_{H}}{4\pi}~,
\end{eqnarray}
\begin{eqnarray}\label{chemical potential}
\mu=\sqrt{\frac{2p}{p-1}}c^{1/2}l~r_{H}~.
\end{eqnarray}

In order to investigate the perturbation in the bulk spacetime and
the critical phenomena when the black hole approaches marginally
stable from the high temperature phase, we will consider the
minimally coupled, charged scalar perturbation,
$\psi_{\varpi,~q}(u)e^{-i(\omega t+kx)}$, with mass $m$, which obeys
the wave equation
\begin{equation}\label{perturbation equation}
\left[u^{p}\frac{d}{du}(\frac{f}{u^{p}}\frac{d}{du})+\frac{J^{2p/(p-1)}(\varpi+\aleph)^{2}}{f(1+c)^{2/(1-p)}}
-\frac{q^{2}}{(1+c)^{2/(1-p)}}-\frac{l^{2}m^{2}J^{2/(p-1)}}{u^{2}}\right]\psi_{\varpi,~q}(u)=0~,
\end{equation}
where we have defined four dimensionless quantities:
\begin{eqnarray}
\varpi&:=&\omega/r_{H}~,\nonumber\\
q&:=&|k|/r_{H}\nonumber~,\nonumber\\
\aleph&:=&\frac{eA_{t}}{r_{H}}=\sigma\left(\frac{1}{1+c}-\frac{u^{p-1}}{J}\right)~,\nonumber\\
\sigma&:=&(1+c)^{p/(p-1)}\frac{e\mu}{r_{H}}.
\end{eqnarray}
Using Eq. (\ref{chemical potential}), we can rewrite $\sigma$ as
\begin{equation}\label{sigma}
\sigma=\sqrt{\frac{2p}{p-1}}~(le)~c^{1/2}(1+c)^{p/(p-1)}~.
\end{equation}
Taking $e\rightarrow\infty$ and keeping $\sigma$ fixed, we can
employ the probe approximation following \cite{maeda}. Equation
(\ref{sigma}) tells that  in the limit
$c\propto(le)^{-2}\rightarrow0$ the background (\ref{new metric})
becomes a neutral Gauss-Bonnet-AdS black hole in $D$
dimensions~\cite{Cai}.

Near the AdS boundary $u\sim0$, Eq. (\ref{perturbation equation})
becomes
\begin{equation}
\left[u^p\partial_{u}\left(\frac{1-\sqrt{1-4\alpha}}{2\alpha}u^{-p}~\partial_{u}\right)-l^{2}m^{2}u^{-2}\right]\psi_{\varpi,~q}(u)=0,
\end{equation}
and $\psi_{\varpi,~q}(u)$ has a fall-off behavior as
\begin{equation}
\psi_{\varpi,~q}(u)\sim\psi_{\varpi,~q}^{-}u^{\lambda_{-}}+\psi_{\varpi,~q}^{+}u^{\lambda_{+}},
\end{equation}
where
\begin{eqnarray}
\lambda_{\pm}:=\frac{1}{2}\left[p+1\pm\sqrt{(p+1)^{2}+4m^{2}l_{eff}^{2}/l^{2}}\right]~.
\end{eqnarray}

In the AdS/CFT duality, the order parameter expectation value
$\langle\mathcal{O}_{\varpi,~q}\rangle$ corresponds to
$\psi_{\varpi,~q}^{+}$ while the source term is
$\psi_{\varpi,~q}^{-}$, so the response function can be defined
by~\cite{maeda}
\begin{equation}\label{response function}
\chi_{\varpi,~q}:=\left.\frac{\delta\langle\mathcal{O}_{\varpi,~q}\rangle}{\delta\psi_{\varpi,~q}^{-}}\right
|_{\psi_{\varpi,~q}^{-}\rightarrow0}\propto\frac{\psi_{\varpi,~q}^{+}}{\psi_{\varpi,~q}^{-}}~.
\end{equation}
We aim to investigate the perturbation and the critical phenomenon,
so that we have to solve the equation of motion of the scalar field,
Eq. (\ref{perturbation equation}), based on the boundary conditions
at the horizon and at the boundary. After obtaining the coefficients
$\psi_{\varpi,~q}^{\pm}$, we can study the behavior of the response
function.

Near the horizon $u\sim1$, Eq. (\ref{perturbation equation}) becomes
\begin{equation}\label{new perturbation equation}
f\frac{\partial}{\partial u}\left[f\frac{\partial}{\partial
u}\psi_{\varpi,~q}(u)\right]+(1+c)^{\frac{2p+2}{p-1}}\varpi^{2}\psi_{\varpi,~q}(u)=0~,
\end{equation}
and its solution is given by $\psi_{\varpi,~q}(u)\sim(1-u)^{\pm
i\frac{\omega}{4\pi T}}$. We impose the ``incoming wave" boundary
condition at the horizon, so $\psi_{\varpi,~q}(u)\sim(1-u)^{-
i\frac{\omega}{4\pi T}}$. Introducing a new variable $\varphi$ as
$\psi_{\varpi,~q}(u)=\mathcal{\Re}(u)\varphi_{\varpi,~q}(u)$ and
choosing
$\mathcal{\Re}(u)=exp[i(1+c)^{1/(p-1)}\int^{u}_{0}du\frac{J^{p/((p-1)}}{f}(\varpi+\aleph)]$,
we can express the boundary condition at the horizon as
$\varphi_{\varpi,~q}(~u=1)=const.$, and Eq. (\ref{perturbation
equation}) becomes
\begin{eqnarray} \label{main equation}
&&\left\{\frac{d^{2}}{du^{2}}+\left[(\frac{d}{du}\ln\frac{f}{u^{p}})+2i\frac{(\varpi+\aleph)
J^{p/(p-1)}}{f(1+c)^{1/(1-p)}}\right]\frac{d}{du}\right.\nonumber\\&&\left.-\frac{q^{2}(1+c)^{2/(p-1)}u^{2}+l^2m^2J^{2/(p-1)}}{u^{2}f}
+i\frac{u^p}{f}\frac{d}{du}\left[\frac{J^{p/(p-1)}(\varpi+\aleph)}{u^{p}(1+c)^{1/(1-p)}}\right]\right\}\varphi_{\varpi,~q}(u)=0~.
\end{eqnarray}

Near the AdS boundary $u\sim0$, $\varphi_{\varpi,~q}$ behaves as
\begin{equation}\label{asymptotic behavior}
\varphi_{\varpi,~q}(u)\sim\varphi_{\varpi,~q}^{-}u^{\lambda_{-}}+\varphi_{\varpi,~q}^{+}u^{\lambda_{+}}~.
\end{equation}
The boundary conditions at the horizon are now given by
\begin{eqnarray}\label{boundary}
\varphi_{\varpi,~q}|_{~u=1}&=&1~,\nonumber\\
\left.\frac{\varphi^{\prime}_{\varpi,~q}}{\varphi_{\varpi,~q}}\right|_{~u=1}
&=&\left.\frac{q^{2}(1+c)^{2/(p-1)}+l^2m^2J^{2/(p-1)}u^{-2} -i
u^p\frac{d}{du}[\frac{J^{p/(p-1)}(\varpi+\aleph)}{u^p(1+c)^{1/(1-p)}}]}{\partial_{u}f-p~
u^{-1}f+2i(\varpi+\aleph)(1+c)^{1/(p-1)}J^{p/(p-1)}}\right|_{~u=1}~.
\end{eqnarray}
Eq. (\ref{main equation}) is a linear equation and
$\varphi_{\varpi,~q}(u)$ must be regular at the horizon. Since we do
not concentrate on the amplitude of $\varphi_{\varpi,~q}(u)$, we can
set $\varphi_{\varpi,~q}(~u=1)=1$.

\section{Numerical Results}

In this section, we will numerically solve Eq. (\ref{main equation})
under the boundary conditions (\ref{boundary}) in the probe
approximation. We will first examine the behavior of the charged
scalar field perturbation, which can present us an objective picture
on how the black hole approaches the marginally stable mode when the
temperature drops. In addition we will determine the coefficients
$\varphi_{\varpi,~q}^{\pm}$ from the asymptotic behavior
(\ref{asymptotic behavior}), and then study the response function
$\chi_{\varpi,~q}\propto\varphi_{\varpi,~q}^{+}/\varphi_{\varpi,~q}^{-}$.
This can tell us the critical behavior of some physical quantities
when the system approaches the critical point. Without loss of
generality, hereafter we will set $e=1$ and AdS radius $l=1$ in our
calculation.

To disclose the high curvature influence on the perturbation and the critical phenomenon, we will concentrate on the 5-dimensional (p=3)
Gauss-Bonnet AdS black holes with the scalar mass $m^2=-3$ and the Gauss-Bonnet coupling parameter within the range
$-\frac{7}{36}\leq\alpha\leq\frac{1}{4}$.

\begin{figure}[ht]
\includegraphics[width=300pt]{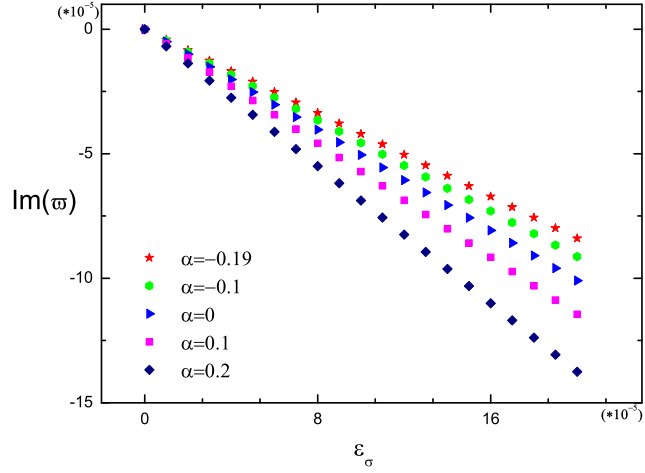}
\caption{\label{fig1}(Color online) The trajectories of the
imaginary parts of the lowest quasinormal frequency for different
values of $\alpha$. While the temperature \textbf{drops to the
critical} point, the system approaches the marginally stable mode.}
\end{figure}

At first we report the influence of the Gauss-Bonnet term on the
scalar perturbation behavior. We concentrate on the lowest
quasinormal frequency which gives the relaxation time \cite{wang}.
We can obtain the quasinormal frequencies by solving Eq. (\ref{main
equation}) based on the boundary conditions (\ref{boundary}) at the
horizon and $\varphi_{\varpi,~q=0}^{-}=0$ at the AdS boundary. The
objective influence of the Gauss-Bonnet term on the imaginary parts
of the lowest quasinormal frequency of the perturbation is shown in
Fig. 1. We see that all the imaginary parts of the quasinormal
frequencies are negative, which shows that the black hole spacetime
is stable. For the larger Gauss-Bonnet coefficient, the imaginary
part of the lowest quasinormal frequency has larger deviation from
zero. This implies that the higher curvature correction can ensure
the system to be more stable and can slow down the process to make
the high temperature black hole phase become marginally stable. This
objective picture of studying the quasinormal modes is consistent
with the observation in \cite{Gregory,Pan-Wang} that the higher
curvature correction can hinder the condensation of the scalar hair
on the boundary. With the decrease of the black hole temperature, we
observe that the lowest quasinormal frequency approaches the origin
and vanishes when the temperature of the system reaches the critical
value, which indicates that the system approaches marginally stable.
The lowest quasinormal frequency approaches the origin with equal
spacing and we fit the results for different Gauss-Bonnet
coefficient in polynomials as below
\begin{eqnarray}
\alpha&=&-0.19,~\varpi_{QNM} \sim (2.57-0.69i)\times10^{-12}
+(1.74-0.42i)~\varepsilon_{\sigma} -
(0.27+0.49i)~\varepsilon_{\sigma}^{2},\nonumber\\
\alpha&=&-0.1,~~~\varpi_{QNM} \sim (1.21-1.50i)\times10^{-13} +
(1.96-0.46i)~\varepsilon_{\sigma} -
(0.37+0.60i)~\varepsilon_{\sigma}^{2},\nonumber\\
\alpha&=&0,~~~~~~~~\varpi_{QNM} \sim (-2.92-0.81i)\times10^{-13} +
(2.23-0.50i)~\varepsilon_{\sigma} -
(0.52+0.76i)~\varepsilon_{\sigma}^{2},\nonumber\\
\alpha&=&0.1,~~~~~\varpi_{QNM} \sim (-1.05+0.23i)\times10^{-9} +
(2.56-0.57i)~\varepsilon_{\sigma} -
(0.72+1.00i)~\varepsilon_{\sigma}^{2},\nonumber\\
\alpha&=&0.2,~~~~~\varpi_{QNM} \sim (1.01-3.40i)\times10^{-13}
+(2.97-0.69i)~\varepsilon_{\sigma} -
(1.28+1.37i)~\varepsilon_{\sigma}^{2},
\end{eqnarray}
where $\varepsilon_{\sigma}=1-\sigma/\sigma_{c}$.

\begin{figure}[ht]
\includegraphics[width=300pt]{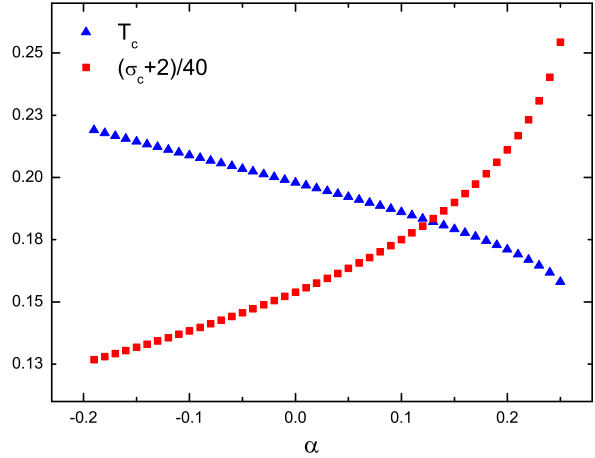}
\caption{\label{fig2}(Color online) The values of
$(\sigma_{c}+2)/40$ (black line) and the critical temperature
$T_{c}$ (blue line) as a function of the Gauss-Bonnet coefficient
$\alpha$. It is shown that the critical temperature decreases with
the increase of the Gauss-Bonnet coefficient.}
\end{figure}

Let's turn to discussing the thermodynamic susceptibility, which can be obtained numerically by solving Eq. (\ref{main equation}) with
$\varpi=q=0$ under the boundary condition (\ref{boundary}) at the horizon and $\varphi_{\varpi=0,~q=0}^{-}=0$ at the AdS boundary using the
shooting method. The dimensionless parameter $\sigma$ determines the phase structure and its critical value $\sigma_{c}$ can be calculated
numerically. In Fig. 2 we exhibit the critical point $\sigma_{c}$ and the critical temperature $T_{c}$ for the background system with different
Gauss-Bonnet coefficient. In the $5$-dimensional spacetime $T_{c}\propto\rho^{1/3}$. The behavior of $T_c$ is consistent with the result obtained
from the analysis of the condensation in \cite{Gregory, Pan-Wang}, which decreases with the increase of the Gauss-Bonnet coefficient. When the
Gauss-Bonnet term disappears, our result goes back to that got in \cite{maeda} for the $5$-dimensional RN-AdS background.

To examine the critical behavior of the thermodynamical
susceptibility $\chi$, we deviate $\sigma$ away from the critical
value $\sigma_c$ and denote the deviation by $\varepsilon_{\sigma}$.
After examining $\varphi^{\pm}_{\varpi=0,~q=0}$ as the function of
$\sigma$ near the critical point for different Gauss-Bonnet
coefficient, we find as expected that the critical behavior
$\varphi^{-}_{\varpi=0,~q=0}$ vanishes while
$\varphi^{+}_{\varpi=0,~q=0}$ approaches to a constant when
$\sigma=\sigma_{c}$. The Gauss-Bonnet term affects the tendency to
the critical behavior which can be observed from the thermodynamical
susceptibility for the stationary homogeneous source
$\chi=\chi_{\varpi=0,~q=0}\propto\frac{\varphi^{+}_{\varpi=0,~q=0}}{\varphi^{-}_{\varpi=0,~q=0}}$
as shown in Fig. 3. The results are fitted by polynomials as below
\begin{eqnarray}
\alpha&=&-0.19,~~~~~~~1/\chi \sim 1.04\times10^{-9} +
0.44~\varepsilon_{\sigma} - 0.59~\varepsilon_{\sigma}^{2},\nonumber\\
\alpha&=&-0.1,~~~~~~~~1/\chi \sim 3.44\times10^{-9} + 0.33
~\varepsilon_{\sigma} -1.19~\varepsilon_{\sigma}^{2},\nonumber\\
\alpha&=&0,~~~~~~~~~~~~~1/\chi \sim 1.74\times10^{-8} +
0.25~\varepsilon_{\sigma} - 3.36~\varepsilon_{\sigma}^{2},\nonumber\\
\alpha&=&0.1,~~~~~~~~~~1/\chi \sim 3.20\times10^{-10} + 0.19
~\varepsilon_{\sigma} - 7.30~\varepsilon_{\sigma}^{2},\nonumber\\
\alpha&=&0.2,~~~~~~~~~~1/\chi \sim 6.20\times10^{-11} +
0.15~\varepsilon_{\sigma} +0.75 ~\varepsilon_{\sigma}^{2}.
\end{eqnarray}
At the critical point $\sigma_c$, $\chi$ diverges as $\chi\propto
1/\varepsilon_{\sigma}$ regardless of the values of the Gauss-Bonnet
coefficient. Defining $\chi\propto|\varepsilon_{\sigma}|^{-\gamma}$,
we find the critical exponent of the thermodynamic susceptibility
$\gamma=1$ at the critical point. Although the Gauss-Bonnet term
cannot modify the critical exponent of the thermodynamic
susceptibility, it does influence the tendency of the thermodynamic
susceptibility when the critical point is approached. In the
vicinity of the critical point, we see that higher curvature
correction has bigger thermodynamical susceptibility.

\begin{figure}[h]
\includegraphics[width=300pt]{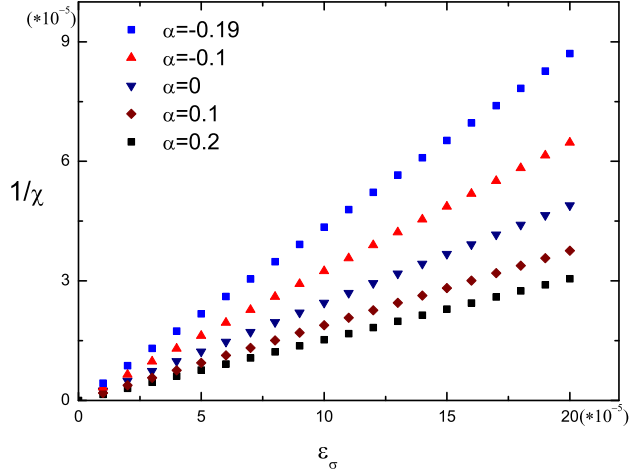}
\caption{\label{fig3}(Color online) The thermodynamic susceptibility
$\chi$ as a function of $\varepsilon_{\sigma}$ for different
$\alpha$. Plotted are $1/\chi$ for the deviation
$\varepsilon_{\sigma}=10^{-5}n$, $n=1,2,\cdots,20$.}
\end{figure}

\begin{figure}[h]
\includegraphics[width=300pt]{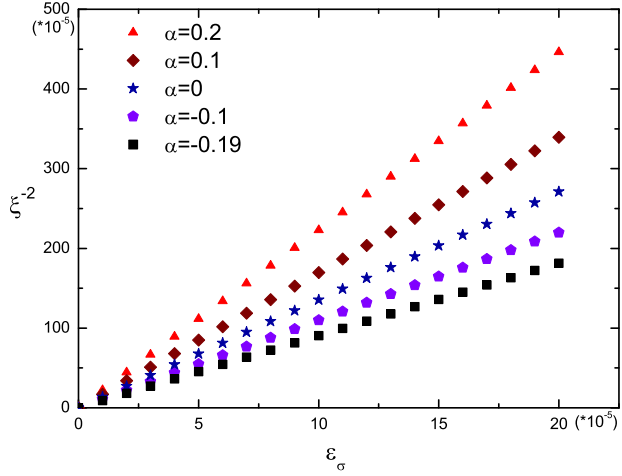}
\caption{\label{fig4}(Color online) The correlation length $\xi$ as
a function of $\varepsilon_{\sigma}$ for different $\alpha$. Plotted
are $\xi^{-2}$ for the deviation $\varepsilon_{\sigma}=10^{-5}n$.}
\end{figure}

Now we shift our gear to discuss the correlation length and static
susceptibility. The critical behavior of the system near the
critical point is determined by the large-scale fluctuations. The
correlation length is the scale parameter that exists in the system
near the phase transition point. It increases while the temperature
approaches its critical value and becomes infinite at the moment of
the phase transition. Now we check the influence imposed by the
Gauss-Bonnet term on the correlation length $\xi$, which is defined
by $\xi^{2}:=-q^{-2}$ \cite{maeda}. We consider a perturbation with
$\varpi=0$ for different $\alpha$, and solve Eq. (\ref{main
equation}) with $\varpi=0$ under boundary conditions: Eq.
(\ref{boundary}) at the horizon and $\varphi_{\varpi=0,~q}^{-}=0$ at
the AdS boundary. Fig. \ref{fig4} shows $\xi^{2}$ with the interval
$\Delta\varepsilon_{\sigma}=10^{-5}$ towards the critical value
$\sigma_{c}$. The results can be fitted by polynomials as
\begin{eqnarray}
\alpha&=&-0.19,~~~~~~\xi^{-2} \sim 1.40\times10^{-11} +
9.07~\varepsilon_{\sigma} -
6.90~\varepsilon_{\sigma}^{2},\nonumber\\
\alpha&=&-0.1,~~~~~~~~\xi^{-2} \sim 8.36\times10^{-12} + 10.99
~\varepsilon_{\sigma} -
8.69~\varepsilon_{\sigma}^{2},\nonumber\\
\alpha&=&0,~~~~~~~~~~~~~\xi^{-2} \sim 1.53\times10^{-12} +
13.55~\varepsilon_{\sigma} -
11.16~\varepsilon_{\sigma}^{2},\nonumber\\
\alpha&=&0.1,~~~~~~~~~~\xi^{-2} \sim 2.05\times10^{-12} + 16.97
~\varepsilon_{\sigma} -
14.50~\varepsilon_{\sigma}^{2},\nonumber\\
\alpha&=&0.2,~~~~~~~~~~\xi^{-2} \sim 2.79\times10^{-12} +
22.32~\varepsilon_{\sigma} - 19.63 ~\varepsilon_{\sigma}^{2}.
\end{eqnarray}
This shows that the correlation length $\xi$ depends on the
Gauss-Bonnet coefficient $\alpha$. For the smaller $\alpha$, the
correlation length is bigger for the same deviation from the
critical point of the system, which means that it is easier for the
system to approach the phase transition point when the Gauss-Bonnet
coefficient is smaller. However at the critical point
$\xi^{-2}\propto\varepsilon_{\sigma}$, i.e.,
$\xi\propto\varepsilon_{\sigma}^{-1/2}$ is always true for all
chosen Gauss-Bonnet coefficients $\alpha$, which shows that the
critical exponent $\nu=1/2$ is independent of the Gauss-Bonnet term.

\begin{figure}[h]
\includegraphics[width=300pt]{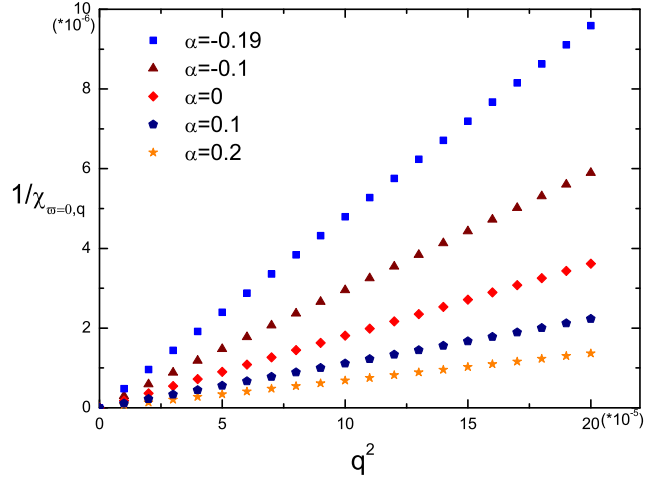}
\caption{\label{fig5}(Color online) The static susceptibility at the
critical point $\chi_{\varpi=0,~q}\mid_{T_{c}}$ as a function of
$q^{2}$ for various $\alpha$. Plotted are
$1/\chi_{\varpi=0,~q}\mid_{T_{c}}$ for $q^{2}=10^{-5}n$.}
\end{figure}

The static critical exponent $\eta$ is determined by the static
susceptibility at the critical point
$\chi_{\varpi=0,~q}\mid_{~T_{c}}\propto q^{\eta-2}$ \cite{maeda} and
can be obtained from its $q$-dependence. Solving Eq. (\ref{main
equation}) with $\varpi=0$ under the boundary condition
(\ref{boundary}) at the horizon, we can get
$\left.\chi_{\varpi=0,~q}\right|_{~T_{c}}\propto~\frac{\varphi_{\varpi=0,~q}^{+}}{\varphi_{\varpi=0,~q}^{-}}$
from the behavior at the AdS boundary. Fig. \ref{fig5} shows
$\left.\chi_{\varpi=0,~q}\right|_{~T_{c}}$ as a function of $q$ and
the fitting results are listed below
\begin{eqnarray}
\alpha&=&-0.19,~~~~~~\left.1/\chi_{\varpi=0,~q}\right|_{~T_{c}} \sim
-1.41\times10^{-11} + 0.048~q^{2} - 0.27~q^{4},\nonumber\\
\alpha&=&-0.1,~~~~~~~~\left.1/\chi_{\varpi=0,~q}\right|_{~T_{c}}
\sim
-2.38\times10^{-10} + 0.030 ~q^{2} -0.65~q^{4},\nonumber\\
\alpha&=&0,~~~~~~~~~~~~\left.1/\chi_{\varpi=0,~q}\right|_{~T_{c}}
\sim 1.92\times10^{-14} + 0.018~q^{2} +0.016~q^{4},\nonumber\\
\alpha&=&0.1,~~~~~~~~~~\left.1/\chi_{\varpi=0,~q}\right|_{~T_{c}}
\sim 2.66\times10^{-14} + 0.011 ~q^{2} +0.027~q^{4},\nonumber\\
\alpha&=&0.2,~~~~~~~~~~\left.1/\chi_{\varpi=0,~q}\right|_{~T_{c}}
\sim 8.96\times10^{-14} + 0.0068~q^{2} -0.0010 ~q^{4}.
\end{eqnarray}
We see that the Gauss-Bonnet term affects the slope of the inverse
static susceptibility. From the fitting result we obtain
$\left.1/\chi_{\varpi=0,~q}\right|_{~T_{c}}\propto q^{-2}$, which
suggests that the exponent $\eta=0$ within numerical errors for
various $\alpha$.

\section{Conclusions and discussions}

We investigated the perturbations of charged scalar field in a $5$-dimensional Gauss-Bonnet-AdS black hole background and paid attention to the
effect of the Gauss-Bonnet term on the critical behavior of the system. From the perturbation behavior we obtained the objective picture on how
the high curvature influences the spacetime perturbation and the formation of the scalar hair. Our results from the dynamical perturbation
support that observed in the study of the condensation phenomena \cite{Gregory,Pan-Wang}. The high curvature will slow down the process for the
system with high temperature to approach the marginally stable state and hinder the condensation of the scalar hair. These effects can also be
read from the susceptibility and the correlation length in the process when the system approaches the marginally stable moment.

We also calculated the critical exponents for holographic superconductors when the critical point of the system is approached from the high
temperature phase. We observed that although the Gauss-Bonnet term affects the processes of the systems to approach the critical moments, they do
not change the static critical exponents, namely the static critical exponents still take the mean-field values. This confirmed the conjecture
that the critical exponents are determined by the matter fields in the system and are independent of the gravity sector of the
system~\cite{maeda}. This is mainly due to the fact that the gravity sector just simply provides a background in the high temperature phase
analysis or in the probe approximation, while the matter fields undergo a second-order phase transition (from zero to nonzero
condensation)~\cite{maeda}. Note that without the charged scalar field, the system does not have any critical phenomenon. This shows the role of
the charged scalar field in the phase transition.

\begin{acknowledgments}

RGC and BW thank the organizers and participants for various
discussions during the workshop on ``Dark Energy and Fundamental
Theory" held at Xidi, Anhui, China, May 28-June 6, 2010, supported
by the Special Fund for Theoretical Physics from the National
Natural Science Foundation of China under grant no: 10947203. This
work was partially supported by the National Natural Science
Foundation of China.

\end{acknowledgments}

\end{document}